\documentclass[prb,twocolumn,amsmath,showpacs]{revtex4-1}
\usepackage{graphicx}
\usepackage{hyperref}
\usepackage{epstopdf}
\usepackage{amssymb,amsfonts,amsmath} 
\usepackage{nicefrac}

\begin{document} \title{
Dynamical regimes  of  dissipative  quantum systems}

\author{D.M.\ Kennes$^1$}  
\author{O.\ Kashuba$^{1,2}$}  
\author{V.\ Meden$^1$}

\affiliation{$^1$ Institut f{\"u}r Theorie der Statistischen Physik, RWTH Aachen University 
and JARA---Fundamentals of Future Information
Technology, 52056 Aachen, Germany}

\affiliation{$^2$Institute of Theoretical Physics, Technische Universit\"at Dresden, 01062 Dresden, Germany}

\begin{abstract} 
We reveal several distinct regimes of the relaxation dynamics of a small quantum 
system coupled to an environment within the plane of the dissipation 
strength and the reservoir temperature.
This is achieved by discriminating between coherent dynamics with damped 
oscillatory behavior on all time scales, partially coherent behavior being 
nonmonotonic at intermediate times but monotonic at large ones, and 
purely monotonic incoherent decay. Surprisingly, elevated temperature 
can render the system `more coherent' by inducing a transition from 
the partially coherent to the coherent regime. This provides a 
refined view on the relaxation dynamics of open quantum 
systems.
\end{abstract}

\date{\today} 
\maketitle
When studying the relaxation dynamics of a small
quantum system coupled to a dissipative environment at
low temperatures $T$ one usually encounters coherent dynamics 
which is damped oscillatory at weak
coupling and monotonic incoherent dynamics at stronger
ones. The best studied example 
comprising this generic physics is the ohmic spin boson model (SBM).~\cite{Leggett87,Weiss12}
This model can be visualized by a spin-$\nicefrac{1}{2}$ degree of freedom with  
tunneling between the up and down states as well as coupling to a bosonic 
reservoir. Nonuniversal effects which depend on the details of the reservoir 
band dominate the behavior at short times $t$ up to $\omega_{\rm c}^{-1}$ set by 
the inverse band width. We are interested in the \emph{universal} aspects 
and exclusively consider the limit in which $\omega_{\rm c}$ is 
much larger than any other energy scale of the problem (scaling limit).
It is well established that in the unbiased case (vanishing Zeeman field)
and for coupling $\nicefrac{1}{2} < \alpha <1$ of the spin and 
the reservoir the expectation value $P(t)= \left< \sigma_z(t) \right>$ 
monotonically approaches zero, i.e.\ the dynamics is incoherent.
At large $t$ it can be described by an exponentially 
decaying function, possibly with subdominant corrections. \cite{Egger97,Kennes13a,Wang08,Kashuba13a,Kashuba13b}
In contrast, for $\alpha \ll 1$ and small $T$, $P(t)$ is a damped oscillatory function (coherent dynamics).  
\cite{Leggett87,Weiss12,Wang08,Anders06,Orth10A,Orth13} 
Even at small $\alpha$ raising $T$ will eventually drive the system
into the incoherent regime. \cite{Garg85,Weiss86,Leggett87,Weiss12,Orth10A,Orth13}

We show that the classification into coherent and incoherent behavior---with
the definition of `coherence' given above---must be refined
and provide a more detailed understanding of the dynamics realized in the 
$\alpha$--$T$ plane by discriminating between 
intermediate and long times.\cite{Kennes13a,Kashuba13b}
Our insights form an improved basis for future studies 
on the relaxation dynamics of dissipative quantum systems as 
investigated in condensed matter physics, 
quantum optics, physical chemistry, and quantum information 
science.~\cite{Leggett87,Weiss12} 

\begin{figure}
\centering
\includegraphics[width=\columnwidth]{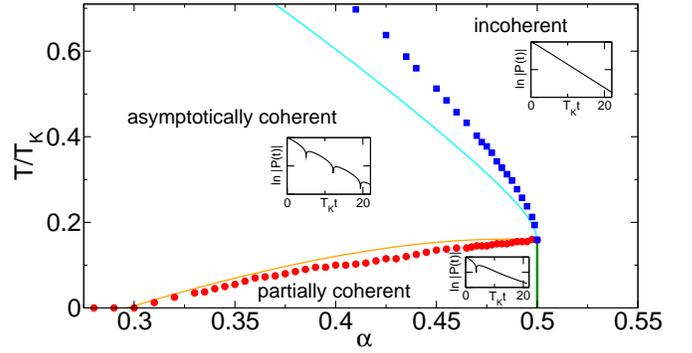}
\caption{(Color online) Diagram showing the extend of the incoherent, asymptotically, 
and partially coherent sectors. The critical temperatures $T_{\rm c1}$ (lower branch) and $T_{\rm c2}$ 
(upper branch) obtained from the numerical solution of the RTRG equations 
\eqref{eq:dpsi} are shown as circles; solid lines correspond to 
the approximatations Eqs.~\eqref{eq:tc1} (lower) and~\eqref{eq:tc2} (upper). 
The insets exemplify the time evolution of the spin expectation value 
$P(t)$ in the different regimes for $\alpha=0.45$ and $T/T_{\rm K}=0.01$ 
(partially coherent), $0.3$ (asymptotically coherent), $0.59$ (incoherent), with
the Kondo scale $T_{\rm K}$.}
\label{fig:phd}
\end{figure}

Our results are summarized in Fig.~\ref{fig:phd} showing the extend of the 
different regimes of distinct dynamical behavior in the $\alpha$--$T$ plane.
For $0< \nicefrac{1}{2}-\alpha \ll 1$ and sufficiently small $T$ we find a regime 
in which $P(t)$ is nonmonotonic (`oscillatory') on intermediate times, but 
monotonic on large ones. In the following we denote this the 
\emph{partially coherent} regime. It must be distinguished from 
the \emph{asymptotically coherent} one encountered at small  
$\alpha$ and $T$,\cite{Leggett87,Weiss12,Garg85,Weiss86} 
in which $P(t)$ shows damped oscillatory behavior on all time scales.
We find a transition line between the partially and the asymptotically 
coherent regimes when raising $T$, which constitutes the central result 
of our work. It implies that for $0<\nicefrac{1}{2}-\alpha \ll 1$ the dynamics 
is \emph{more coherent at elevated temperatures} than at low ones, which 
is rather counterintuitive. At even larger $T$, $P(t)$ will eventually 
become purely monotonic and the dynamics is \emph{incoherent}. For $\nicefrac{1}{2} < \alpha < 1$ 
it is incoherent for all $T$.  

\begin{figure}
\centering
\includegraphics[width=1.01\columnwidth]{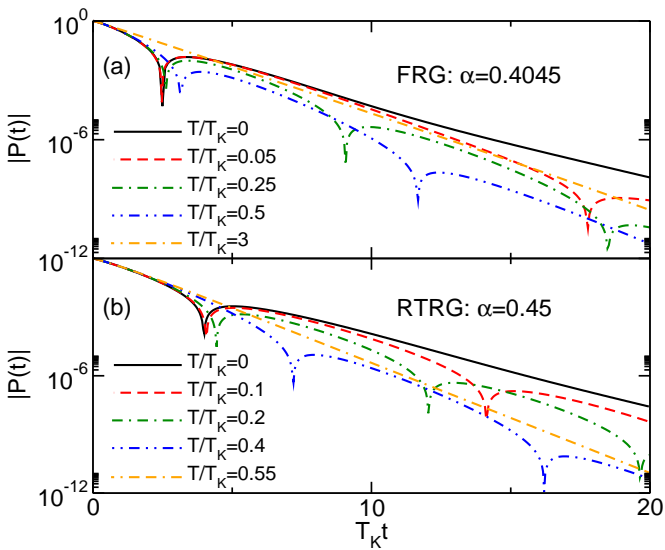}
\caption{(Color online) Time dependence of the spin expectation value $|P(t)|$ on a linear-logarithmic 
scale [(a) FRG, (b) RTRG]. Both approaches capture the transition from partially coherent to 
asymptotically coherent to incoherent dynamics when raising $T$.}
\label{fig:pt}
\end{figure}

At $T=0$ the appearance of the partially coherent regime in between the more 
standard asymptotically coherent and incoherent ones can be shown 
analytically.\cite{Egger97,Kennes13a,Kashuba13b} For $T>0$ our numerical results 
for $P(t)$ obtained by two complementary renormalization group (RG) approaches---the 
real-time RG (RTRG)~\cite{Schoeller09} and the functional RG (FRG) \cite{Metzner12}---indicate 
the transition between the partially and the asymptotically coherent regimes when increasing $T$. 
In Fig.~\ref{fig:pt} we show $|P(t)|$ for different $T$ at fixed $\alpha$ close to 
$\nicefrac{1}{2}$ (see also the insets of Fig.~\ref{fig:phd}). 
At $T=0$ one starts out in the partially coherent regime with $P(t)$ having a single zero 
(dip in a linear-logarithmic plot of $|P(t)|$). Increasing $T$, more zeros appear signalling the transition
to the asymptotically coherent regime; next, the distance between the
zeros increases and eventually all of them disappear when entering 
the incoherent one.  

\begin{figure}[t]
\centering
\includegraphics[width=0.85\columnwidth]{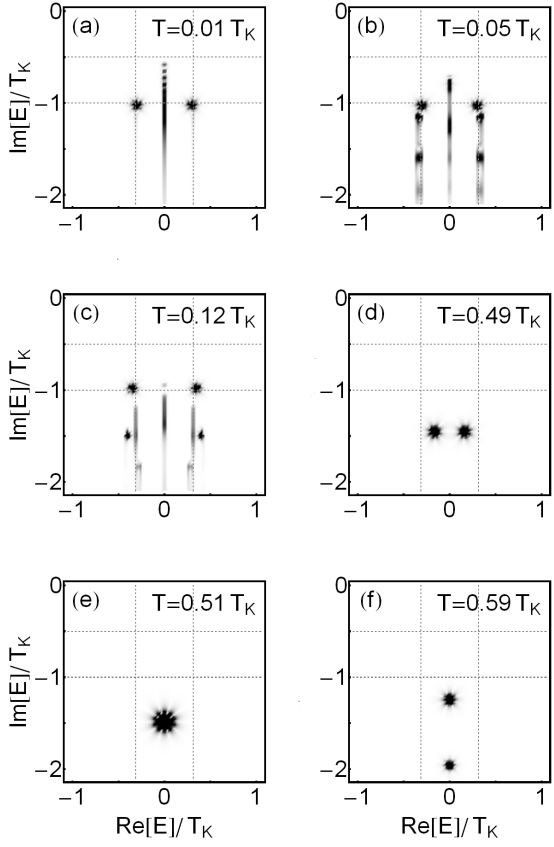}
\caption{Nonanalytical features of the propagator $\Pi_{1}(E)$ at coupling $\alpha=0.45$.
Density plots of $|\partial_{\mathrm{Im}\,E}\mathrm{Re}\,\Pi_{1}+\partial_{\mathrm{Re}\,E}\mathrm{Im}\,\Pi_{1}|$ 
as a function of the complex variable $E$ are shown. Dashed horizontal lines indicate the zero temperature decay 
rates for the poles $T_{\rm K}$ and branching point $T_{\rm K}/2$. Vertical ones indicate the frequency $\pm \Omega$ of the 
poles at $T=0$ ($\Omega \approx \pi g T_{\rm K}$ 
.\cite{Egger97,Kennes13a,Kashuba13b,Lesage98})}
\label{fig:cauchy}
\end{figure}

A more detailed understanding can be obtained from studying the 
Laplace transform $\Pi_1(E) = \int_0^\infty dt \, e^{i E t} P(t)$ of the spin
expectation value in the lower half of 
the complex $E$-plane. 
This also allows to precisely determine the transition termperatures.
Our RTRG approach is set up in Liouville-Laplace 
space~\cite{Pletyukhov12,Kennes13a,Kashuba13b,Kashuba13a} and $\Pi_1(E)$  can be accessed directly through  
the numerical integration of the RG flow equations. The relaxation dynamics is determined by the nonanalyticities 
of the propagator $\Pi_1(E)$ at $z_n \in {\mathbb C}$. Each yields a 
separate contribution to $P(t)$ of the form $\sim \exp{(i z_n t)}$, 
where the frequency and decay rate are determined by $\mbox{Re} \, z_n $ and 
$\mbox{Im} \, z_n$, respectively. 
For branch cuts the exponential decay is accompanied by weakly $t$-dependent 
corrections. \cite{Egger97,Kennes13a,Kashuba13b,Wang08,Kashuba13a} 
The  nonanalyticity closest to the real axis dominates the dynamics at large $t$.
Figure \ref{fig:cauchy} highlights the singularities and branch cuts (dark features) of  $\Pi_1(E)$  for 
$\alpha=0.45$ and different $T$.
For $0< \nicefrac{1}{2}-\alpha \ll 1$ and $T=0$, $\Pi_1(E)$ has a pair of poles with nonvanishing 
real parts of equal absolute value [damped oscillations of $P(t)$] as well 
as a branch cut on the imaginary axis [monotonic decay of $P(t)$]. The distance of the start 
of the branch cut to the real axis is smaller than the distance between the latter 
and the poles. Thus for large $t$ the branch cut term prevails giving rise to partially coherent 
relaxation dynamics.\cite{Kennes13a,Kashuba13b} For $T>0$ the branch cut disintegrates into
singularities, \cite{Garg85,Weiss86} as can be seen in Fig.~\ref{fig:cauchy}(a).  
Raising $T$ these move down while the imaginary part of the finite 
frequency poles  barely changes, see Fig.~\ref{fig:cauchy}(b).
When the top zero frequency singularity passes the level of the pole pair 
$T_{\rm c1}(\alpha)$ is reached and the system undergoes a transition into the 
asymptotically coherent regime, see Fig.~\ref{fig:cauchy}(c).
Further incerasing $T$ the oscillation frequency decreases as the pair of 
poles starts to move inwards, while the singularities on the imaginary axis 
continue to move down and leave the frame shown in Fig.~\ref{fig:cauchy}, 
see Fig.~\ref{fig:cauchy}(d). The frequency vanishes when the 
pole pair hits the imaginary axis at $T_{\rm c2}(\alpha)$, indicating the transition from 
asymptotically coherent to incoherent dynamics, see Fig.~\ref{fig:cauchy}(e).
At further increasing $T$ two singularities move in opposite directions along the 
imaginary axis such that the one moving up approaches the 
decay rate of $\alpha=\nicefrac{1}{2}$ for $T \to \infty$, see Fig.~\ref{fig:cauchy}(f). 
The $T_{\rm c 1/2}$ obtained by such an analysis varying $T$ \emph{and} $\alpha$ are 
shown as circles in Fig.~\ref{fig:phd}. 
The additional finite frequency features (black circles with gray tails) visible 
in  Figs.~\ref{fig:cauchy}(b) and (c) are artifacts of our leading order 
(in $\nicefrac{1}{2}-\alpha$) approximation (see below).

{\it Model and setup}---The unbiased SBM is given by the Hamiltonian 
\begin{eqnarray}
H = - \frac{\Delta}{2} \sigma_x + \sum_k \omega_k b_k^\dag b_k^{} - \sum_k
 \frac{\lambda_k}{2} \sigma_z \left(  b_k^\dag + b_k^{}\right),  
\end{eqnarray}
with the Pauli matrices $\sigma_\nu$, $\nu=x,z$, bosonic ladder operators $b_k^{(\dag)}$, 
tunneling amplitude $\Delta$, reservoir dispersion $\omega_k$, and coupling $\lambda_k$.
The spin-boson coupling is characterized by a spectral density $J(\omega) = \sum_k \lambda_k^2 
\delta(\omega-\omega_k)$. Its $\omega$ dependence is set by the details of the microscopic 
model underlying the SBM.\cite{Leggett87} We focus on the ohmic case with 
$J(\omega)= 2 \alpha \omega \Theta(\omega_{\rm c} -\omega)$, $\alpha \geq 0$; 
$\omega_{\rm c}$ and $\Delta$ only enter in the combination 
$T_{\rm K} =\Delta\left(\Delta / \omega_{\rm c}\right)^{\alpha/(1-\alpha)}$ 
defining the emergent Kondo scale.\cite{Leggett87,Weiss12} 

We prepare the system as the product of the spin-up state (in $z$-direction) 
and the canonical density matrix ($T>0$) for the reservoir.
At time $t=0$ the coupling between the spin and the bosons is switched on 
and the relaxation sets in. For $\alpha<1$ and infinitely large times 
a steady state with $\lim_{t \to \infty} P(t)=0$ is reached.
Here we do not consider the quantum phase transition to the localized regime 
with $\lim_{t \to \infty} P(t) \neq 0$ at $\alpha=1$.\cite{Leggett87,Weiss12}
We considered other initial density matrices and verified 
that the classification of the dynamics is not affected by the particular 
choice of the $t=0$  state.

Investigating the most interesting regime $|\nicefrac{1}{2}-\alpha| \ll 1$ 
we do not study the SBM but rather 
employ the mapping to the interacting resonant level model 
(IRLM).\cite{Leggett87,Weiss12,Kashuba13b} Within the IRLM 
$\alpha=\nicefrac{1}{2}$ corresponds to the noninteracting limit. Our 
RG methods provide controlled access to the relaxation dynamics around this 
exactly solvable point. For the FRG this was earlier shown for  $T\geq 0$.\cite{Kennes12,Kennes13b} In fact, the RG 
flow equations for the one-particle irreducible vertex functions of  
Ref.~\onlinecite{Kennes13b} can directly be used to numerically compute $P(t)$. 
For $T=0$ RTRG equations were derived in Refs.~\onlinecite{Kennes13a,Kashuba13b}. 

{\it RTRG for finite temperatures}---We next discuss how to extend 
the RTRG approach to $T>0$ and describe the steps to derive analytical expressions
for $T_{\rm c1/2}(\alpha)$. 
Readers interested in results only can skip this part. 

In the RTRG approach one aims at the reduced density matrix of 
the spin.\cite{Schoeller09,Pletyukhov12} 
Integrating out the reservoirs' degrees of freedom at $T>0$ leads to a 
summation over Matsubara frequencies $\omega_{m}=\pi T(2m+1)$. The RG equations 
for the relaxation rates $\Gamma_{1/2}$ derived in Refs.~\onlinecite{Kennes13a,Kashuba13b} 
for $T=0$ thus change to
\begin{eqnarray}
\frac{d\Gamma_{1/2}(E)}{dE} & = & i g \Gamma_1(E) \, 2\pi T 
\sum_{m=0}^{\infty}\Bigl[\Pi_{2/1}\bigl(E+i\omega_{m}\bigr)\Bigr]^{2},
\label{eq:deqsum}
\\ \Pi_{n}(E) & = & i\left[E+i\Gamma_{n}(E)/n\right]^{-1}, \;\; n=1,2
\end{eqnarray}
where $g=1-2\alpha$ is the small parameter and the initial conditions read 
$\Gamma_{1/2}(i\omega_{\rm c})=\Delta^{2}/\omega_{\rm c}$.
To perform the Matsubara sum to order $g$ we neglect the $E$-dependence 
of $\Gamma_{1/2}$ in $\Pi_{2/1}$, which yields
\begin{equation}
\frac{d\Gamma_{1/2}(E)}{dE} = i g \frac{\Gamma_1(E)}{2\pi T}
\psi'\left(\frac{1}{2}+\frac{1}{2\pi T \Pi_{2/1}(E)}\right) ,
\label{eq:dpsi}
\end{equation}
where $\psi'(z)$ is the trigamma function. This set of equations was solved 
numerically to obtain Fig.~\ref{fig:cauchy} and, after an additional (numerical) 
Laplace back transform, Fig.~\ref{fig:pt}(b).  

The right hand side of the differential equation \eqref{eq:dpsi} contains 
a series of second order poles, which after integration turn 
into essential singularities of the propagator $\Pi_1$ at $z_n$.
They result from  the disintegration of the branch cut of 
$\Gamma_{1}(E)$ found at $T=0$ \cite{Egger97,Kennes13a,Kashuba13b} and are located on the 
imaginary axis \cite{Garg85,Weiss86} [see Fig.~\ref{fig:cauchy}(a)]; their mutual distance 
is $2 \pi T[1+{\mathcal O}(g)]$.
Close to those singularities we find $\Pi_{1}\propto e^{-g/(E-z_n)}$.
The finite frequency branch cuts (black circles with gray tails) of  
Figs.~\ref{fig:cauchy}(b) and (c) are 
artifacts of the lowest order truncation. In the full diagrammatic series underlying 
the RTRG approach \cite{Schoeller09} no integrals over the reservoirs excitation frequencies appear for $T>0$, but only summations over $\omega_{m}$. Therefore, branch cuts are excluded.
The contributions of the artificial nonanalyticities to $P(t)$, however, are of order $g^{2}$. In 
addition, they are always located below the leading order singularities and do not affect $P(t)$ at 
long times.

We next present the steps to derive analytical approximations for $T_{\rm c1/2} (\alpha)$ 
(curved lines in Fig.~\ref{fig:phd}). 
Solving Eq.~(\ref{eq:dpsi}) for $\Gamma_1$ to leading order in $g$ one can neglect the weak $E$-dependence 
of $\Gamma_{2}$ and finds    
\begin{align}
\Gamma_{1}(E) &= \widetilde{T}_{\rm K} e^{- g \psi \left(\frac{1}{2}+\frac{-iE+\Gamma_{2}(E)/2}{2\pi T}\right) },
\label{eq:gamma1}
\end{align}
where $\widetilde{T}_{\rm K}=T_{\rm K}\left(2\pi T/T_{\rm K}\right)^{-g}$ results from the high-energy 
integration limit and $\psi$ denotes the digamma function.
The positions of the nonanalyticities of $\Pi_1(E)$ can now be determined approximately 
using Eq.~\eqref{eq:gamma1}.
The finite frequency poles $z_{\pm}=\pm\Omega-i\Gamma_{1}^{*}$ are the solutions of 
$iz_{\pm}=\Gamma_{1}(z_{\pm})$,
where we can additionally set 
$\Gamma_{2}(E) \approx \Gamma_{1}(E)$. At $T=0$ this was reasoned to be a good 
approximation.\cite{Kennes13a,Kashuba13b} It yields the self-consistancy equation 
$\Gamma_{2}^{*}=\Gamma_{2}(z_{0})$ with $z_{0}=-i\pi T-i\Gamma_{2}^{*}/2$.
Approximating $\Gamma_{2}(E)$ by  $\Gamma_2^\ast$ in Eq.~\eqref{eq:gamma1} and 
using  Eq.~\eqref{eq:dpsi} we obtain $\Gamma_{2}^{*}=2\pi T g\int_{0}^{\infty}u(x)\psi'(x+u(x)-\Gamma_{2}^{*}/4\pi T)dx$, 
where $u(x)=(T_{\rm K}/2\pi T)^{1-g}e^{-g\psi(x)}$. 

The \emph{first transition} occurs at $\Gamma_{1}^{*}=\Gamma_{2}^{*}/2+\pi T_{\rm c1}$. For 
decreasing $\alpha$, $T_{\rm c1}(\alpha)$ 
decreases and we replace the digamma function in Eq.~\eqref{eq:gamma1} by its low-temperature 
($2\pi T\ll T_{\rm K}$) expansion  $\psi(x)\approx\log x$ ($|x|\gg 1$). Equation \eqref{eq:gamma1}  then 
simplifies to $\Gamma_{1}(E)=T_{\rm K}\left[(-iE+\pi T+\Gamma_{2}(E)/2)/T_{\rm K}\right]^{-g}$,
which is the $T=0$ result \cite{Kennes13a,Kashuba13b} with a $\pi T$ shift on the right hand side.
With this the rates $\Gamma_{1/2}^\ast$ are to leading order given by the $T=0$ expressions 
\cite{Kennes13a,Kashuba13b}
\begin{equation}
\label{eq:rates}
\frac{\Gamma_1^{\ast(0)}}{T_{\rm K}} = \mbox{Im} \, e^{(i\pi+\ln 2)\frac{g}{1+g}} , \; 
\frac{\Gamma_2^{\ast(0)}}{T_{\rm K}} = 2 \left[\frac{\pi g}{2 \sin (\pi g)} \right]^{\frac{1}{1+g}} .
\end{equation}
The \emph{second transition} takes place if   
$\Gamma_{1}^{*}=\Gamma_{1}(-i\Gamma_{1}^{*})$ has a single real valued solution $\Gamma_{1}^{*}$ 
(collapse of finite frequency poles). For decreasing $\alpha$, $T_{\rm c2}(\alpha)$  increases, and 
we apply $\psi(x)\approx-\gamma-1/x$  ($|x|\ll 1$) for $2\pi T\gg T_{\rm K}$ in Eq.~\eqref{eq:gamma1};
$\gamma$ denotes the Euler constant.
The critical temperatures $T_{\rm c1/2}(\alpha)$ resulting from these approximations are given in 
the next section. 

For $\alpha \to \nicefrac{1}{2}-0^+$, where $\Gamma^*_{1/2}=T_K$, $T_{\rm c1/2}$ can be calculated without approximating 
the digamma function in Eq.~(\ref{eq:gamma1}). With
$\Gamma^*_1=\Gamma^*_2/2+\pi T_{{\rm c}1}$ we find $T_{{\rm c}1}(\nicefrac{1}{2})=T_K/(2\pi)$. For 
$T_{{\rm c}2}(\nicefrac{1}{2})$ we aim at a single real valued solution of $\Gamma_1(-i \Gamma^*_1)=\Gamma^*_1$, 
which implies $d\Gamma_1(-i \Gamma^*_1)/d\Gamma^*_1=1$. Using Eq.~\eqref{eq:gamma1} with 
$\Gamma_1(E)\approx \Gamma_2(E)$ leads to 
$d\Gamma_1(-i \Gamma^*_1)/d\Gamma^*_1=g\Gamma^*_1\psi'([\pi T-\Gamma^*_1/2]/[2 \pi T])=4\pi T$.
For $g \to 0$ this equation can only be fulfilled for vanishing argument of the 
trigamma function, which gives $T_{{\rm c}2}(\nicefrac{1}{2})=T_K/(2\pi)$.

{\it Results}---The physics obtained from the numerical solution of our RTRG and FRG \cite{Kennes12,Kennes13b}
flow equations---which both are controlled for $|\nicefrac{1}{2}-\alpha| \ll 1$---was already discussed in the 
first section; we here add further details. 

The crucial element of our reasoning is
the analytical structure of the Laplace transform $\Pi_{1}(E)$ of the spin expectation value $P(t)$
in the lower half of the complex $E$ plane. To single out the nonanalyticities in Fig.~\ref{fig:cauchy}, 
we use the Cauchy-Riemann relations and show 
$|\partial_{\,\mathrm{Im}\,E}\mathrm{Re}\,\Pi_{1}(E)+\partial_{\,\mathrm{Re}\,E}\mathrm{Im}\,\Pi_{1}(E)|$ 
as a function of $E$.  
If we would be able to solve the RTRG equations analytically---or in the impractical limit of an infinitely 
dense grid in the numerical solution---this expression would diverge at the singularities and branch cuts and would be zero elsewhere. For a numerical solution on a finite grid it becomes a very efficient 
tool for highlighting the areas in the vicinity of nonanalyticities. As described above from 
plots of this type $T_{\rm c1/2}$ can be extracted. 

At $\alpha = \nicefrac{1}{2}$ we find a  `triple point' 
with $T_{\rm c1}(\nicefrac{1}{2})=T_{\rm c2}(\nicefrac{1}{2})=T_{\rm K}/(2\pi)$ (for analytical results, see the 
last section). The noninteracting blip approximation (NIBA) \cite{Leggett87,Weiss12} 
only captures the transition line $T_{\rm c2}(\alpha)$ to the incoherent regime. Within 
this method one obtains $\lim_{\alpha \to \nicefrac{1}{2}-0^+} T_{\rm c2}= T_{\rm K}/\pi$.\cite{Garg85,Weiss86} 
In NIBA the singularities of $\Pi_{1}(E)$ located on the imaginary axis are treated incorrectly. 
This implies that $\Gamma_2^\ast/2$ is absent in the equation defining  $T_{\rm c2}(\nicefrac{1}{2})$  
which can be linked to the missing factor $\nicefrac{1}{2}$.  U.~Weiss informed us that using improved 
NIBA \cite{Egger97} at finite $T$ gives $T_{\rm c2}(\nicefrac{1}{2})$ in agreement with our 
result.\cite{Weiss13} 

The approximate analytical solution of the RTRG equations 
provides us with expressions for $T_{\rm c1/2}$ also away from $\alpha =\nicefrac{1}{2}$. We find
\begin{equation}
T_{\rm c1}(\alpha)=\frac{1}{\pi}\left(\Gamma_{1}^{*(0)}-\Gamma_{2}^{*(0)}/2\right),
\label{eq:tc1}
\end{equation}
with $\Gamma_{1/2}^{*(0)}$ of Eq.~(\ref{eq:rates}) and $g=1-2\alpha$ (lower curved line in Fig.~\ref{fig:phd}). 
This yields a very good approximation to the numerically obtained $T_{\rm c1}$ (circles in  Fig.~\ref{fig:phd}) 
for $0.3 < \alpha <\nicefrac{1}{2}$. 
At the lower bound $T_{\rm c1}$ vanishes. We can thus estimate the critical coupling separating the
partially and asymptotically coherent regimes at $T=0$ by $\alpha_{\rm c} \approx 0.3$. Strictly speaking 
such $\alpha$'s are beyond the regime $|\nicefrac{1}{2}-\alpha| \ll 1$ in which our approximate 
RTRG and FRG equations (derived for the IRLM) are 
controlled. However, the numerical solution of the $T=0$ RTRG equations for 
the \emph{SBM at weak coupling} $\alpha \ll 1$,\cite{Kashuba13a} which is complementary to the present approach,
confirms the asymptotically to partially coherent transition 
and gives $\alpha_{\rm c} \approx 0.36$. This indicates that our results can be trusted 
even down to $\alpha \approx 0.3$ and that the exact $\alpha_{\rm c}$ is located close to this value.      
The second transition temperature (upper curved line in Fig.~\ref{fig:phd}) can be approximated as 
\begin{equation}
T_{\rm c2}(\alpha)=\frac{T_{\rm K}}{2\pi}e^{\frac{g(1+\gamma)+\sqrt{2g+g^{2}}}{1+g}} \!\! \left(1 \! + \! g \! 
+ \! \sqrt{2g \! + \! g^2}\right)^{\frac{1}{1+g}},
\label{eq:tc2}
\end{equation}
where  $\gamma$ is the Euler constant.
For $\alpha$ close to $\nicefrac{1}{2}$ ($g \ll 1$) this simplifies to 
\begin{equation}
T_{\rm c2}(\alpha) \approx\frac{T_{\rm K}}{2\pi}\left(1+4\sqrt{\frac{1}{2}-\alpha}\right).
\end{equation}

{\it Summary}---We investigated the relaxation dynamics of the ohmic spin-boson model---the prototype 
model of dissipative quantum mechanics---as a function of temperature and dissipation strength. 
We identified a regime in which the spin expectation value is nonmonotonic at short to 
intermediate times but monotonic at large ones, the partially coherent regime. For spin-boson 
couplings $0.3 \lessapprox \alpha < \nicefrac{1}{2}$ the dynamics for $0 \leq T < T_{\rm c1}(\alpha)$ is only partially 
coherent while for $T_{\rm c1}(\alpha) \leq T < T_{\rm c2}(\alpha)$ it is asymptotically 
coherent, that is (damped) oscillations appear on all time scales. In contrast to the general expectation   
that larger $T$ will foster dissipation and thus suppress coherence we find that elevated temperature 
enhances coherence. Only for $T > T_{\rm c2}(\alpha)$ the system enters the incoherent regime described 
earlier.\cite{Garg85,Weiss86} 

\begin{acknowledgments}
We thank M.~Pletyukhov, H.~Schoeller, and U.~Weiss for discussions 
and the DFG for  support (FOR 723).
\end{acknowledgments}

{}

\end{document}